\documentstyle[12pt]{article}
\begin{document}


\title{CAN LIGO SEE COMPACT BINARIES?}
\author{Serge Droz\footnote{e-mail: droz@physics.uoguelph.ca}
\mbox{} and Eric Poisson\footnote{e-mail: poisson@physics.uoguelph.ca}\\
 Dept. of Physics,
University of Guelph, Guelph, ON \\N1G 2W1, CANADA}

\maketitle

\begin{abstract}
The probability that interferometric detectors such as LIGO and VIRGO will
successfully detect inspiraling compact binaries 
depends in part on our knowledge of the expected gravitational wave forms. 

The best approximations to the true wave forms available today are the 
post-Newtonian (PN) templates. In this paper we argue that these 2PN templates
are accurate enough for a successful search
for compact binaries with the advanced LIGO interferometer.
Results are presented for the  40-meter Caltech prototype as well as for the
inital and advance LIGO detectors.
\end{abstract}

\section{INTRODUCTION}

Currently (November 1997) there are five large scale interferometric 
gravitational-wave detectors under construction, namely the two American LIGO detectors, 
the French Italian VIRGO, the German-British GEO600 and finally the Japanese
TAMA. The immediate goal of this effort is the direct detection of gravitational
waves. In a later stage this international network of detectors will act as
a gravitational wave observatory with a variety of scientific projects (see
reference \cite{Thorne_K:1997} for more details). The most promising sources
for detection are inspiraling binaries of black holes and/or neutron stars.

Because since gravitational waves are so weak, any signal will likely be obscured
by detector noise. The method of choice to search for a {\em known} signal in a
noisy data stream is matched filtering \cite{Wainstein_Z:1962}:
The data is compared to a set of template wave forms for a ``best
fit'' (a more precise definition is given below). Matched filtering is an
optimal method in the sense that no other (linear) filtering method is more
effective. The drawback is that a set of very accurate templates is needed to
track the signal with the necessary accuracy. For example a typical
inspiraling-binary signal might
approximately 15000 wave cycles in the advanced LIGO frequency band. This
means that we need a template that traces all of the 15000 cycles. By missing only one cycle the
detection probability falls drastically. Unfortunately, in general relativity
there is no way to (analytically) calculate the exact wave-form and we have to
resort to some kind of approximation. The most promising approximative
templates are obtained by a post-Newtonian expansion of the gravitational waves
in the orbital velocity of the binary companions.

In this contribution we  assess the quality of the PN templates by comparing
them to signals obtained from black hole perturbation theory. This is
known to provide very accurate wave forms when the binary mass ratio is small.

\section{MATCHED FILTERING}

Let us write the detector output as
\[
s(t) = h(t) + n(t),
\]
where $n(t)$ denotes the detector noise and $h(t)$ a possibly absent signal.
We define the squared signal to noise ratio (SNR) by
\begin{equation}
  \rho^2 = \frac{(s,t)}{{\rm rms}(n,t)}.
  \label{rho} 
\end{equation}
Here the inner product $(\  ,\  )$ is defined by
\begin{equation}
  (s, t) := 2 \int_0^\infty \frac{df}{S_n(f)} 
     \left(\hat{s}(f)\hat{t}^*(f) + \hat{s}^*(f)\hat{t}(f) \right),
  \label{eq:scalarp}
\end{equation}
where a hat $\hat{}$ denotes a Fourier transform and an asterisk $^*$ 
complex conjugation. $S_n$ is the (one sided) spectral density of the noise.
It is large where the noise is large. Consequently, regions of low noise get a higher
weight in the integral (\ref{eq:scalarp}), the ``match'' between the template $t$
and the detector output $h$. If in an actual measurement $\rho$ exceeds a
certain threshold then detection is claimed.

Wieners theorem \cite{Wainstein_Z:1962} now states that $\rho$ is maximal if $t =
h$ in equation (\ref{rho}). This choice of template is called the Wiener
optimal filter. Conversely, if $t \neq h$ then we get a smaller
$\rho$ and thus might miss signals that otherwise would have been detected.

Since we want to make statements about the templates for ``typical'' noise, we
usually talk about the the {\em expectation value} of the SNR. 
Under the assumption of Gaussian noise it can be calculate to be \cite{Cutler_C:1994}
\[
  \langle \rho^2 \rangle = \underbrace{\frac{(h,t)}{\sqrt{(t,t) 
  (h,h)}}}_{\cal A} 
  \langle \rho^2 \rangle_{\rm best} . 
\]
The quantity $\langle \rho^2 \rangle_{\rm best} = \sqrt{( h, h )}$ is 
the SNR one would obtain if one knew the exact waveform $h$. The 
function ${\cal A}$ is called the {\em ambiguity function} and it 
depends on the template used. Maximizing ${\cal A}$ over all templates 
gives the {\em fitting factor} $FF$. This is the reduction of the SNR 
caused by inaccurate templates: $\langle \rho^2 \rangle_{\rm max} = FF 
\langle \rho^2 \rangle_{\rm best}$. The fitting factor therefore 
measures the quality of the templates.

The best templates available today are calculated using a 
post-Newtonian expansion \cite{Blanchet_L:1996} of Einstein's 
equations. For inspiraling binaries, the wave forms are expressed as a power 
series in $v/c$ where $v$ is the orbital velocity. (Recently Pad\'e 
approximates have been  used to improve the convergence of the 
post-Newtonian expansion \cite{Damour_T:1997}).

To simplify matters, one often works with the so-called restricted 
post-Newtonian approximation given by
\begin{equation}
  t(f) \propto f^{-\frac{7}{3}} e^{i \Psi({\cal M},\eta,t_c,\phi_c)},
  \label{pnt} 
\end{equation}
where the amplitude is kept at the lowest post-Newtonian order but the 
phase $\Psi({\cal M},\eta,t_c,\phi_c)$ includes higher PN 
corrections. Here we keep $\Psi$ accurate to 2 PN order.
 This makes sense, because we expect a slight mismatch in the 
phases to lead to a rapidly oscillating integrand in 
(\ref{eq:scalarp}) and thus to a low SNR. Here the  chirp mass ${\cal 
M} = (m_1 m_2)^\frac{3}{5}/(m_1 + m_2)^\frac{1}{5}$, the mass ratio
$\eta = (m_1 m_2)/(m_1 + m_2)^2$, the arrival time $t_c$ and the 
initial phase $\phi_c$
parameterize the templates. 

The test signal used to calculate the fitting factor is obtained 
from black hole perturbation theory \cite{Poisson_E:1993}, 
integrating the 
Teukolsky equation \cite{Teukolsky_S:1973}. The signal is given as an
infinite sum over modes and takes the form \cite{Droz_s:1997}
\begin{equation}
  h(f) \propto \sum_{l=2}^{\infty} \sum_{m=1}^{l} A^{lm}(v) 
  e^{\psi(v)} S_l(\vartheta,\varphi)
  \label{testsignal}
\end{equation}
where $ v = (2 \pi M f/m)^{1/3}$ ($M = m_1 + m_2$). The functions 
$A_{lm}(v)$ and $\psi(v)$ must be calculated numerically and $S_l$ is a known
function.
The $l=m=2$ mode dominates the signal and we have verified 
numerically that modes with $l>3$ can be neglected in the sum 
(\ref{testsignal}).

To calculate the fitting factor one sticks expressions (\ref{pnt}) 
and (\ref{testsignal}) into the definition of the ambiguity function 
and maximizes over the parameters ${\cal M}$, $\eta$, $t_c$ and 
$\phi_c$ for a ``fixed'' signal (\ref{testsignal}).

\section{RESULTS AND INTERPRETATION}

\begin{table}[ht]
\begin{center}
\begin{tabular}{ccccc}
\hline
 & \multicolumn{2}{c}{advanced LIGO} & initial LIGO & 40-meter \\

System & $FF$-Newton & $FF$- 2-PN & $FF$- 2-PN & $FF$- 2-PN \\
\hline \hline

$1.4-1.4 M_\odot$ & $79.2 \% $ & $93.0 \% $ & $95.8\%$ & $94.0 \% $ \\
$0.5-5.0 M_\odot$ & $51.6 \% $ & $95.2 \% $ & $89.7\%$ & $ 89.0 \% $ \\
$1.4-10 M_\odot$ & $55.8 \% $ & $91.4 \% $ & $88.4\%$ & $< 85.0 \% $ \\
$10-10 M_\odot$ & $70.1 \% $ & $91.7 \% $ & $86.3\%$ &  \\
$4-30 M_\odot$ & $61.3 \% $ & $86.6 \% $ & $67.9\%$ &  \\

\hline
\end{tabular}
\end{center}
\caption{The fitting factors for different noise curves and binary 
systems are listed. A fitting factor  larger than $90\%$ is considered 
acceptable. Note that only the $1.4-1.4 M_\odot$ system, for which 
$FF > 90\%$, is within the targeted range of the 40-meter detector. }

\end{table}

We have calculated  the fitting factors for a variety 
of different binaries and detector noise curves (for the initial and advanced LIGO 
detectors as well as the 40m Caltech prototype). In the former 
case, in addition to the 2 PN templates we have also tested the 
Newtonian ($\equiv$ lowest PN) templates. The results are summarized in 
Table 1.

We find that for the mass range of interest to a particular 
detector, the fitting factors for the 2PN templates
are all above $90\%$. This makes them suitable for detection of 
inspiraling binary signal \cite{Apostolatos_A:1996}. 
For small mass ratios the fitting factors  drop considerably 
reflecting the fact that the systems become more relativistic for 
fixed chirp mass ${\cal M}_c$. One would expect PN templates to perform less well 
in this limit. 

An additional point should be made. Formally, black hole perturbation 
theory assumes $\eta \equiv 0$. Thus it is not clear if our 
analysis holds for nearly-equal-mass systems. However, by setting 
$\eta = 0$ in the templates (\ref{pnt}) one effectively obtains a PN 
expansion of the perturbation-theory signal. When we do so we do not find any 
significant changes in the fitting factor. This leads us 
to believe that our results should hold, at least qualitatively,  even for nonzero 
mass ratios. 

In conclusion, we may say that the 2 PN templates are suitable for 
the detection of binary inspiral signals.


\begin{thebibliography}{1}

\bibitem{Thorne_K:1997}
K.~S. Thorne,
\newblock preprint  (1997), gr-qc/9704042.

\bibitem{Wainstein_Z:1962}
L.~A. Wainstein and V.~D. Zubakov,
\newblock  (Prentice-Hall, London, 1962).

\bibitem{Cutler_C:1994}
C.~Cutler and E.~E. Flanagan,
\newblock Phys. Rev. {\bf D49}, 2658 (1994), gr-qc/9402014.

\bibitem{Blanchet_L:1996}
L.~Blanchet, B.~R. Iyer, C.~M. Will, and A.~G. Wiseman,
\newblock Class. Quant. Grav. {\bf 13}, 575 (1996), gr-qc/9602024.

\bibitem{Damour_T:1997}
T.~Damour, B.~R. Iyer, and B.~S. Sathyaprakash,
\newblock preprint  (1997), gr-qc/9708034.

\bibitem{Poisson_E:1993}
E.~Poisson,
\newblock Phys. Rev. {\bf D47}, 1497 (1993).

\bibitem{Teukolsky_S:1973}
S.~A. Teukolsky,
\newblock APJ {\bf 185}, 635 (1973).

\bibitem{Droz_s:1997}
S.~Droz and E.~Poisson,
\newblock Phys. Rev. {\bf D56}, 4449 (1997), gr-qc/9705034.

\bibitem{Apostolatos_A:1996}
T.~A. Apostolatos,
\newblock Phys. Rev. {\bf D54}, 2421 (1996).

\end{thebibliography}

\end{document}